\documentclass[10pt]{article}
\usepackage[a4paper, left=0.75in, right=0.75in, top=1in, bottom=1in]{geometry}
\usepackage{multicol, caption}
\usepackage{amsmath,amssymb}
\usepackage{graphicx}
\usepackage{float}
\usepackage{cite}
\usepackage{multirow}
\usepackage{caption}
\usepackage{siunitx}
\usepackage[version=4]{mhchem}
\usepackage{hyperref}
\usepackage[affil-it]{authblk}  % for affiliation management

\title{\textbf{On the Strategies to Enhance zT}}
\allowdisplaybreaks

\author{D. Beretta\thanks{Corresponding Author: davide.beretta@uzh.ch \\University of Zurich}}
        
\date{\small December 19, 2024}

\begin{document}

\maketitle

\begin{abstract}
Enhancing the dimensionless figure of merit $zT$ is central to developing better thermoelectric materials and advancing thermoelectric generation technology. However, the intrinsic interdependence between electrical conductivity, the Seebeck coefficient, and thermal conductivity presents a significant challenge. Over time, various strategies have emerged, but the literature remains difficult to navigate due to its widespread distribution across numerous sources. This short review highlights the key approaches to improving $zT$, offering a clear and concise guide to help researchers understand the major ideas and breakthroughs in the field.
\end{abstract}

\vspace{2\baselineskip}

\begin{center}
    \begin{minipage}{0.8\textwidth}
        \tableofcontents
    \end{minipage}
\end{center}

\newpage

\begin{multicols}{2}
\section{Introduction}
The efficiency $\eta$ of a TEG is defined as the ratio between the electrical power output $P$ delivered to a load and the heat absorbed from the heat source $Q_{in}$, expressed as $\eta=P/Q_{in}$. Both the power output and the efficiency depend on the \textbf{thermocouple dimensionless figure of merit} $ZT=S_{pn}^2 T / R_{pn} K_{pn}$, where $S_{pn}$, $R_{pn}$, and $K_{pn}$ are the Seebeck coefficient, the electrical resistance series and the thermal conductance parallel of a thermocouple, respectively, and $T$ is the absolute temperature.\cite{Beretta.Caironi.Materials.Science.and.Engineering.R.Reports.2019, Beretta.Caironi.Sustainable.Energy.&.Fuels.2017} The higher the $ZT$, the higher $\eta$ and $P$, making it a critical parameter in the performance of TEGs. 
\begin{figure}[H]
    \includegraphics[width=1\linewidth]{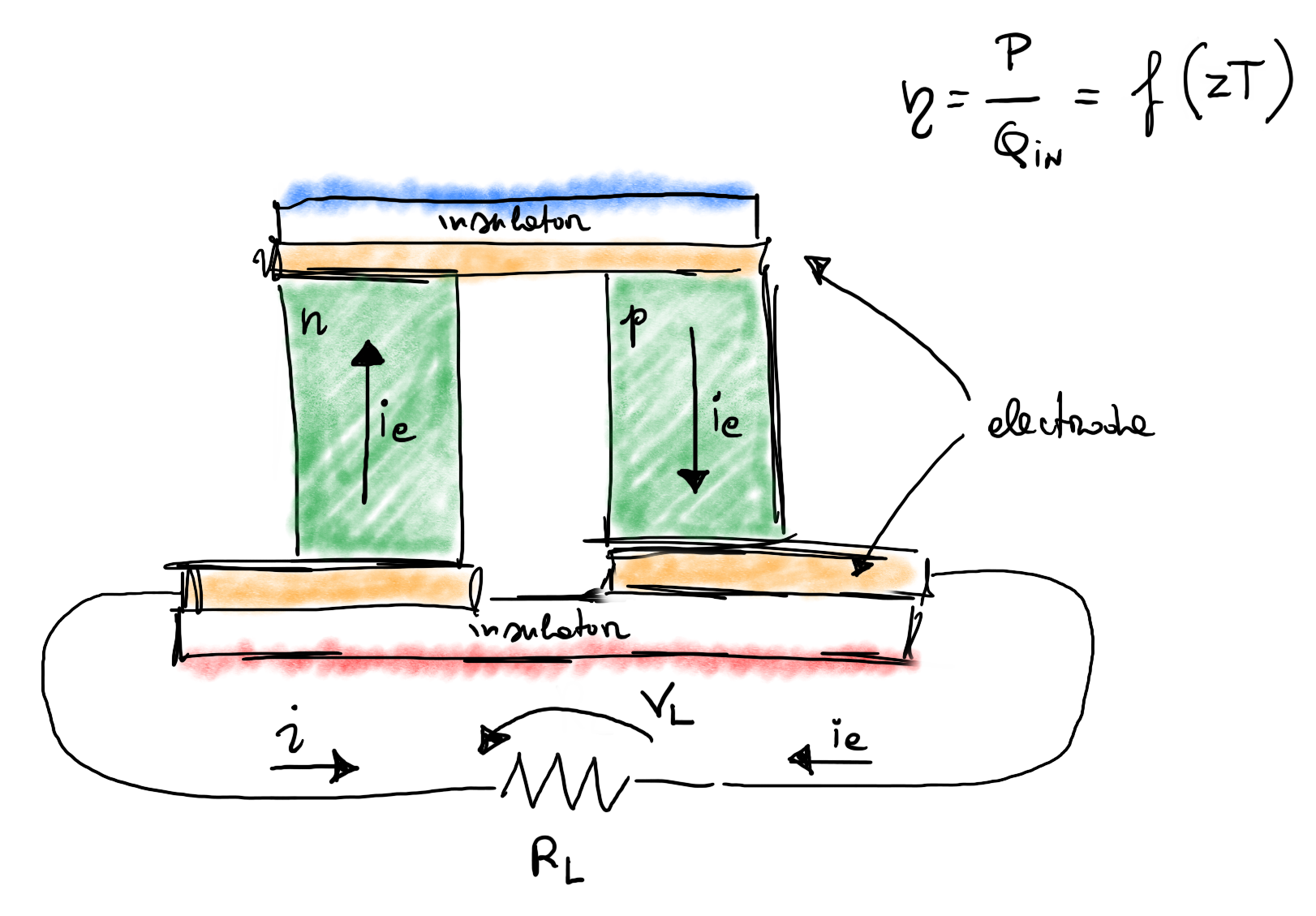}
    \caption{Illustration of a thermoelectric generator consisting of one thermocouple and a load resistance $R_L$. The red and blue shaded areas indicate the hot and cold sides of the device, respectively, while the electrodes are represented in yellow.}
    \label{fig:teg}
\end{figure} \noindent
The ideal thermocouple would therefore behave as a perfect electrical conductor, a perfect thermal insulator, and possess a large Seebeck coefficient.
Inspired by the definition of $ZT$, the thermoelectric quality of a material is similarly characterized by the \textbf{material dimensionless figure of merit} $zT = \sigma S^2 T/ \kappa$, where $\sigma$, $S$ and $\kappa$ are the electrical conductivity, the Seebeck coefficient and the thermal conductivity of the material, respectively, and $T$ is the absolute temperature. $ZT$ (uppercase $Z$) is related to the $zT$ (lowercase $z$) of the p- and n-type materials that form the thermocouple. While their relationship cannot be expressed in closed form,\cite{Ioffe.Infosearch} in general, the higher the $zT$ of the materials, the greater the conversion efficiency and power output of a TEG. 
Notably, in cases where transport is ballistic, the electrical conductance $G$ and thermal conductance $K$ replace the electrical conductivity $\sigma$ and thermal conductivity $\kappa$, leading to the expression $zT = S^2 G T / K$. Regardless of the transport regime, the ideal thermoelectric material would combine perfect thermal insulation, perfect electrical conduction, and a large Seebeck coefficient.
However, the design and optimization of thermoelectric materials are challenging due to the inherent interdependence of the transport coefficients $\sigma$, $S$, and $\kappa$. Achieving high $zT$ has, therefore, necessitated strategies aimed at decoupling these coefficients to enable their independent optimization. Over the years, this has led to the development of several innovative approaches.
This short review aims to summarize these strategies and their most successful demonstrations. While not exhaustive, it provides a concise and accessible guide to help scientists and researchers navigate this vast and complex field, where the interchangeable use of terminology can sometimes lead to confusion.
  
\section{Strategies}
The following subsections provide an overview of key strategies to enhance $zT$, namely doping, lattice disorder, micro- and nano-grains, low-dimensional systems, phononic nanocrystals, phononic metamaterials, nanoscale interfaces, and thermionic generation in heterostructures. While the categorization of these strategies is not unique, as many of the approaches involve overlapping material engineering techniques, this work adopts the following specific criteria for classification: micro- and nano-grains are discussed separately from low-dimensional systems, as the latter focus on the distinct quantum properties of structures such as quantum wells and quantum wires, while the former emphasize nanostructures as components of a macroscopic system. Phononic heterostructures are treated separately from low-dimensional systems because, in the former, transport occurs perpendicular to the growth direction (or parallel to the quantization axis), whereas in the latter, it occurs normal to the quantization direction. Thermionic generation in heterostructures is also addressed separately from phononic nanocrystals, as the primary goal of thermionic generation is energy filtering—rather than reducing lattice thermal conductivity by opening phononic band gaps—and because thermionic generation may not require a periodic heterostructure. Finally, nanoscale interfaces are considered a part of—and the limit case of—low-dimensional systems.

\subsection{Doping}
\begin{figure}[H]
    \centering
    \includegraphics[width=0.9\linewidth]{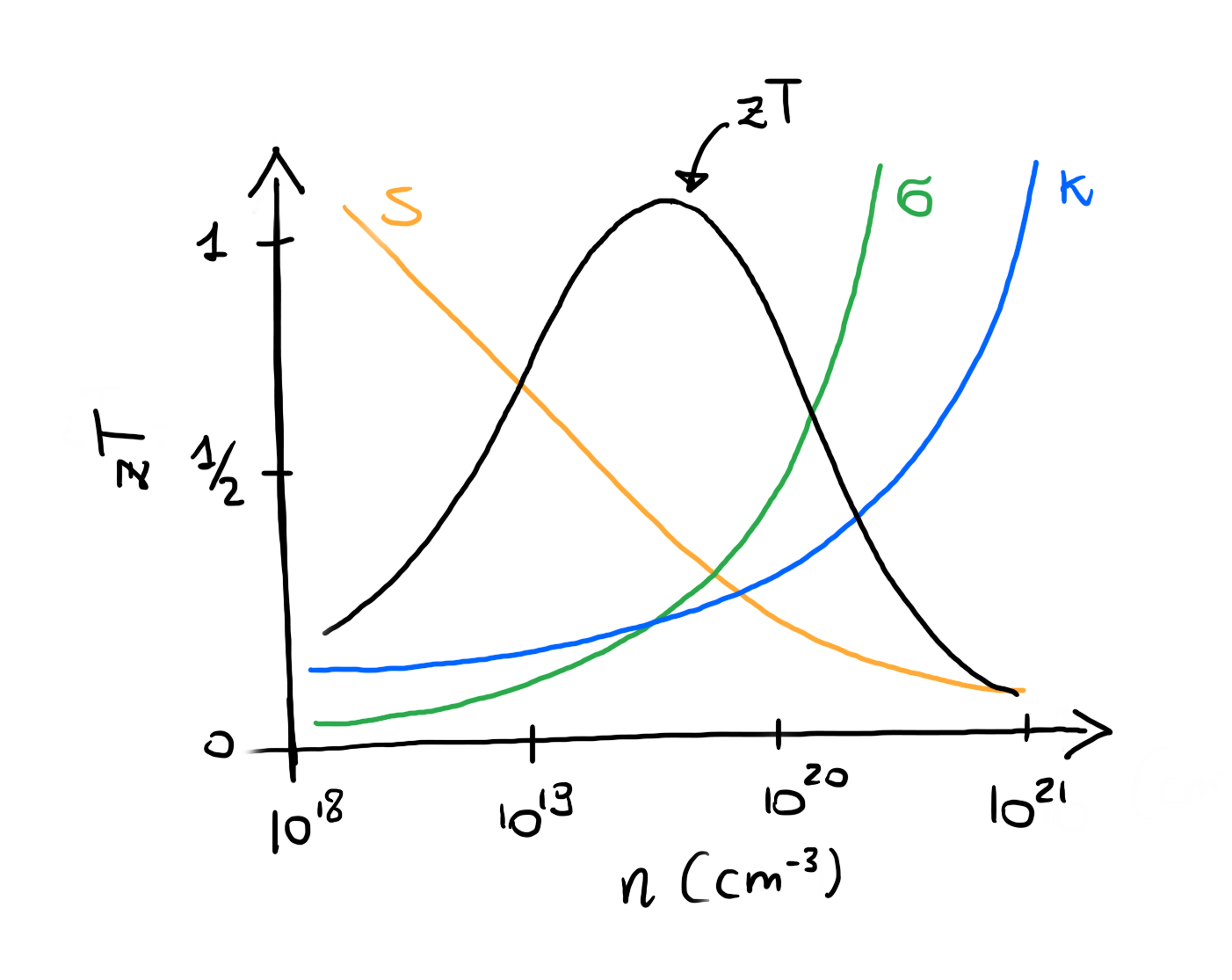}
    \caption{Qualitative plot of $\sigma$, $S$, $\kappa$ and $zT$ as a function of the free charge carrier density $n$. Note that $zT$ is normalized between zero and one. Adapted from Synder et al.\cite{Snyder.Toberer.Nature.Materials.2008}}
    \label{fig:optimal_doping}
\end{figure}

There exists an optimal density of free charge carriers $n$ to maximize the $zT$ of a thermoelectric material. This occurs because $\kappa$ comprises contributions from both the lattice $\kappa_l$ and free charge carriers $\kappa_e$, where $\kappa_e$ depends on $n$; $\sigma$ is a function of $n$; and $S$ depends on the symmetry of the density of states around the Fermi level, whose position is also influenced by $n$. 
Figure~\ref{fig:optimal_doping} qualitatively illustrates the dependence of $S$, $\sigma$, and $\kappa$ on the free charge carrier density. While both $\sigma$ and $\kappa$ increase with $n$, with $\kappa \rightarrow \kappa_l$ at low carrier density, $S$ decreases as $n$ increases. This reduction occurs because only the charge carriers near the Fermi level contribute to the Seebeck coefficient. In particular, electrons above the Fermi level diffuse towards the cold side, while those below diffuse towards the hot side, as determined by the difference between the Fermi distributions at the respective temperatures. At higher carrier densities, the difference in the number of states above and below the Fermi level within a few $k_B T$ becomes less pronounced. This leads to electron currents from hot to cold and from cold to hot with similar magnitudes, which tend to cancel each other out, resulting in a smaller $S$.
Good thermoelectric materials are typically lightly doped semiconductors with $n$ in the range $\SI{1e19}{}-\SI{1e21}{\per\centi\meter\cubed}$, as this density provides the best compromise between the transport coefficients. Consequently, materials for thermoelectric applications are typically doped to achieve a free charge carrier density within this range.
\cite{Snyder.Toberer.Nature.Materials.2008} 

\subsection{Lattice Disorder} \label{sec:lattice_disorder}
\begin{figure}[H]
    \centering
    \includegraphics[width=0.9\linewidth]{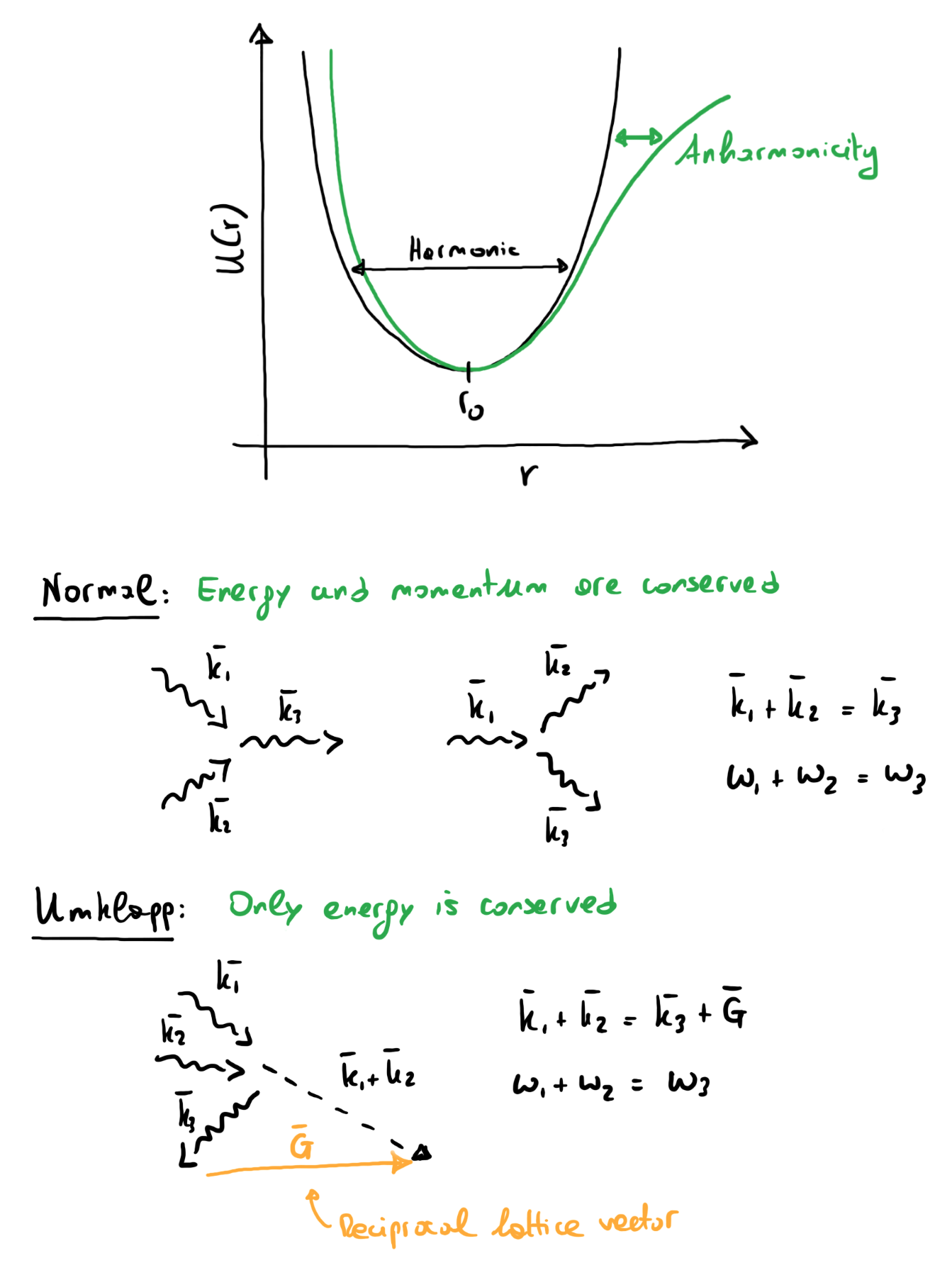}
    \caption{Illustration of Normal and Umklapp phonon-phonon scattering processes. Scattering occurs naturally, also in perfect crystals, as a result of the anharmonic potential. Since Normal scattering conserve momentum, the latter does not contribute to thermal conductivity reduction. However, Umklapp scattering occurs mostly with phonons at the Brillouin zone boundaries. Therefore, Normal process, which redistribute crystal momentum, must be taken into account when modeling thermal conductivity, as they can result in phonon wave vectors close to the Brillouin boundary.}
    \label{fig:normal_umklapp}
\end{figure}

The thermoelectric figure of merit $zT$ can be enhanced by reducing the lattice contribution to thermal conductivity $\kappa_l$ through the introduction of lattice disorder. This disorder can be introduced during material synthesis (or growth) by forming point defects, such as interstitials, vacancies, and substitutional atoms, or by partially filling empty cages in complex crystal structures with rattling atoms \cite{Snyder.Toberer.Nature.Materials.2008}. Lattice disorder enhances anharmonicity in the lattice potential, which increases phonon scattering, thereby reducing the thermal conductivity $\kappa_l$ without significantly altering the electrical conductivity $\sigma$. For a brief review of the main phonon scattering processes, see Fig.~\ref{fig:normal_umklapp}.

\begin{figure}[H]
    \centering
    \includegraphics[width=1\linewidth]{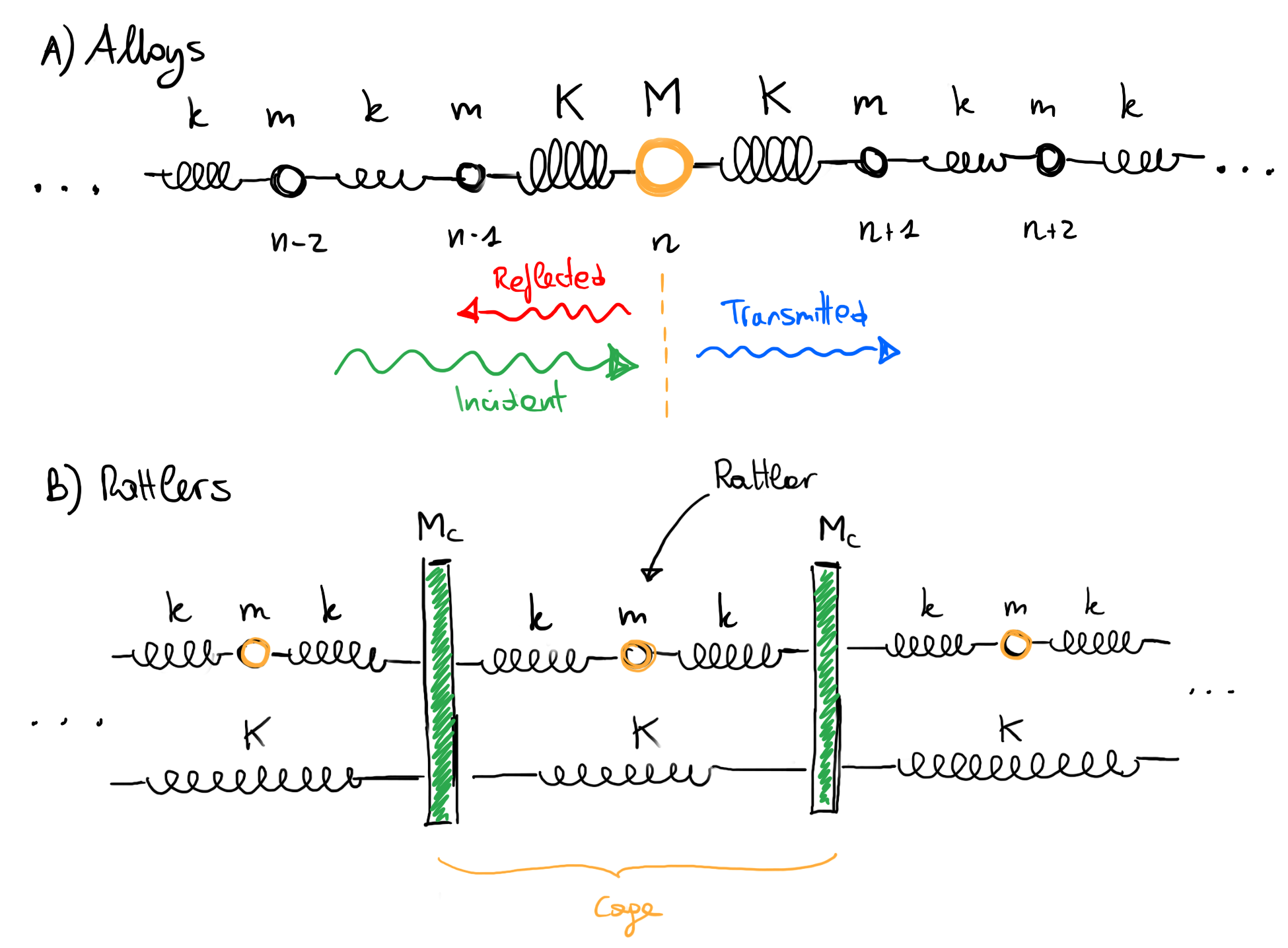}
    \caption{Illustration of a simplified model based on a mechanical spring system to represent lattice disorder caused by (A) alloying and (B) rattling atoms (adapted from Christensen et al.\cite{Christensen.Iversen.Nature.Materials.2008}). Here, $M_c$ and $K$ represent the mass and spring constant of the cage, which consists of several atoms, while $m$ and $k$ are the mass and spring constant of the guest atom.}
    \label{fig:lattice_disorder}
\end{figure}

One of the most effective methods to introduce such disorder is \textbf{alloying}, a technique proposed in the late 1950s. The central idea is that the lattice thermal conductivity $\kappa_l$ of an alloy is lower than that of its pure constituents due to differences in atomic mass and elastic constants between substitutional atoms and host crystal atoms \cite{Ioffe.Infosearch}, while the electrical conductivity $\sigma$ remains relatively unaffected.\cite{Wang.Snyder.Advanced.Functional.Materials.2013} Experimental evidence is abundant, and it is no coincidence that state-of-the-art room-temperature thermoelectric materials are based on alloys such as \ce{Bi2Te3} with \ce{Sb2Te3} (e.g., \ce{Bi_{0.5}Sb_{1.5}Te3}, p-type) and \ce{Bi2Te3} with \ce{Bi2Se3} (e.g., \ce{Bi2Te_{2.7}Se_{0.3}}, n-type), which both achieve $zT \approx 1$ at room temperature. This is, in part, due to a $\kappa_l$ that is roughly half the value of the alloy’s constituents.\cite{Goldsmid.Springer} 
While first-principles methods such as Perturbative Density Functional Theory (PDFT) are required for a detailed understanding of phonon scattering in anharmonic potentials, the underlying physics of lattice disorder due to alloying can be illustrated using a simple model of a linear atomic chain containing a single guest atom, as shown in Fig.~\ref{fig:lattice_disorder}A. 
 \noindent
In this model, the host atoms have the same mass $m$ and elastic constant $k$, while the guest atom has a different mass $M$ and elastic constant $K$. The presence of the guest atom breaks the crystal's translational symmetry, enabling elastic scattering of incident phonons, particularly those at frequencies resonant with the guest atom's vibrational modes. The scattering process is mathematically described by the Fermi’s golden rule and can be visualized as a superposition of reflected and transmitted waves that satisfy the boundary conditions at the guest atom. While elastic scattering dominates, it is important to note that inelastic scattering is also possible, in which incoming phonons lose energy (and momentum), exciting localized vibrational modes on the guest atom. 
A widely used analytical model to quantify the ratio of the alloy’s thermal conductivity to that of the pure material is the Klemens model.\cite{Klemens.Klemens.Proceedings.of.the.Physical.Society.Section.A.1955} In this model, the ratio depends on a disorder parameter that quantifies the variance in mass and elastic constants within the system. Greater disorder leads to stronger phonon scattering, resulting in a further reduction of the thermal conductivity $\kappa_l$.
 \noindent

Another highly effective strategy to reduce lattice thermal conductivity involves using complex crystal structures where \textbf{rattling atoms} occupy large cages within the host material’s lattice. These guest atoms are weakly bound and can undergo large vibrations (rattling) around their equilibrium positions. Similar to alloying, fully describing the physics of rattlers requires quantum mechanical simulations of phonons in lattices with anharmonic potentials. However, the lattice disorder induced by rattlers can be visualized using the simple model illustrated in Fig.~\ref{fig:lattice_disorder}B. Here, the guest atom is connected to its cage (comprising several host atoms), while adjacent cages are coupled to each other and to the guest atom. Even this simplified classical model predicts an energy gap in the phonon dispersion relation.\cite{Christensen.Iversen.Nature.Materials.2008} The size of the energy gap is proportional to the coupling strength between the guest and host atoms, with larger cages and more weakly bound guest atoms resulting in a wider gap. 
The avoided crossing of phonon modes caused by this energy gap reduces the group velocity of phonons near the gap, given by $v_g = \partial \omega / \partial k$, thereby lowering the thermal conductivity. Additionally, phonons with frequencies that fall into the gap are scattered at the guest atom sites due to the absence of allowed vibrational modes in that energy range. This mechanism has been shown to significantly reduce the lattice thermal conductivity in medium- to high-temperature thermoelectric materials, including Clathrates, Skutterudites, and Zintl phases.\cite{Snyder.Toberer.Nature.Materials.2008}

A special case of lattice disorder is found in \textbf{amorphous materials}, which inherently exhibit very low thermal conductivity, \( \kappa \sim \SI{1}{\watt\per\meter\per\kelvin} \), due to the absence of phonon-mediated thermal transport. However, these materials typically suffer from extremely low electrical conductivity, making them unsuitable for thermoelectric applications. Nevertheless, certain organic materials based on doped polymers, such as poly(3,4-ethylenedioxythiophene) polystyrene sulfonate (PEDOT:PSS) and poly(3,4-ethylenedioxythiophene) tosylate (PEDOT:Tos), exhibit a coexistence of crystalline and amorphous phases, enabling them to achieve electrical conductivities as high as \( \SI{100} - \SI{1000}{\siemens\per\centi\meter} \).\cite{Beretta.Caironi.ACS.Applied.Materials.&.Interfaces.2017, Bubnova.Crispin.Nature.Materials.2014} In these materials, the crystalline phase provides efficient pathways for charge transport, while the interplay of crystalline and amorphous regions creates unique electronic properties, such as the semi-metallic behavior observed in PEDOT:Tos. These materials have attracted significant interest for room-temperature thermoelectrics, with PEDOT:Tos demonstrating $zT \approx 0.5$ at room temperature, attributed to its semi-metallic nature.\cite{Bubnova.Crispin.Nature.Materials.2014} Unfortunately, these materials are highly sensitive to environmental conditions, posing significant challenges to their potential practical applications.

\subsection{Micro- and Nano-Grains}\label{sec:micro_nano_grains}
\begin{figure}[H]
    \centering
    \includegraphics[width=0.9\linewidth]{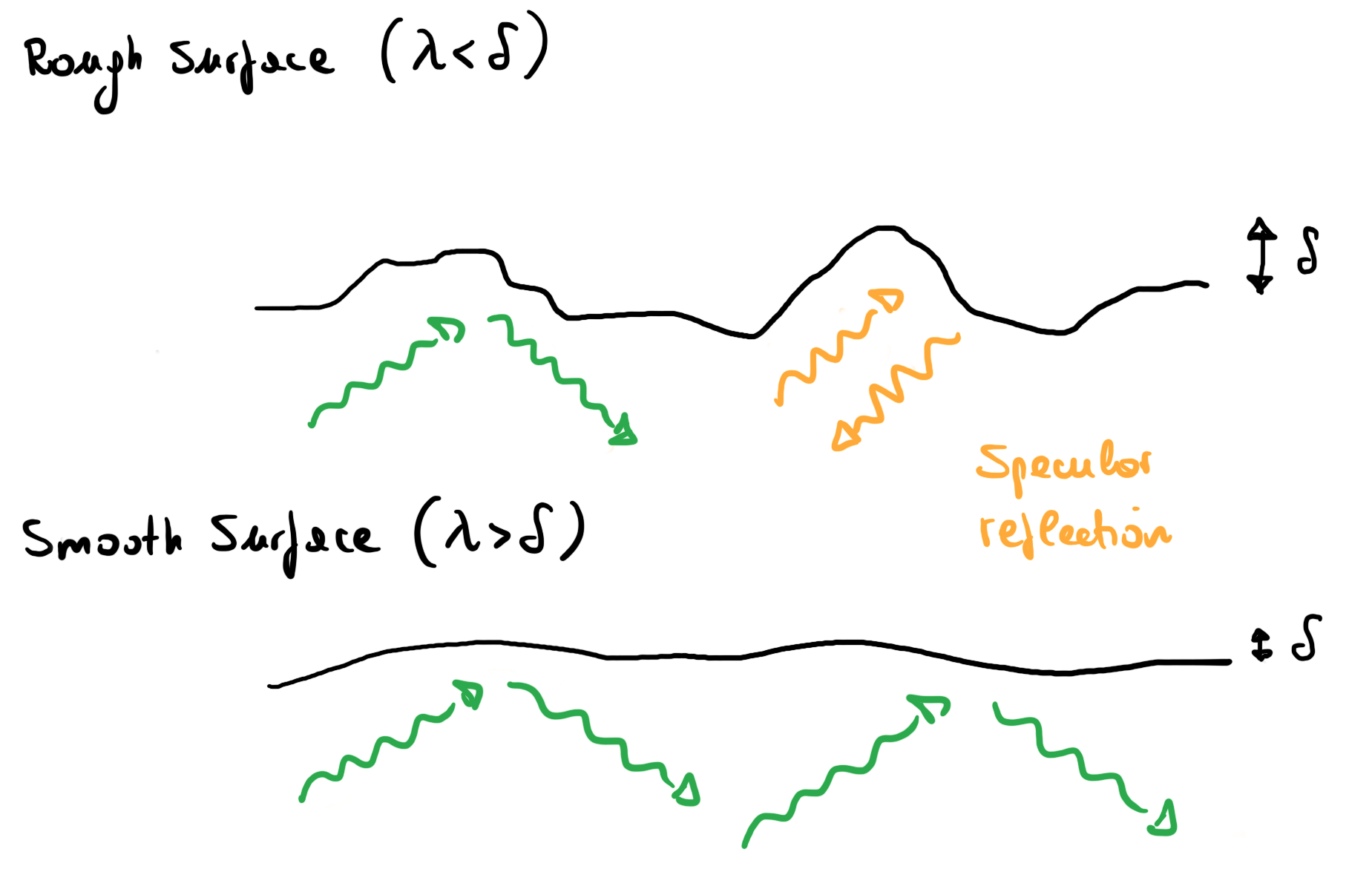}
    \caption{Surface roughness $\delta$. Rough surfaces are prone to reflect heat waves more than smooth surfaces.}
    \label{fig:surface_roughness}
\end{figure}

The phonon wavelength spectrum contributing to $\kappa_l$ is broad, typically ranging from a few \SI{}{\nano\meter} up to a few \SI{}{\micro\meter}, depending on the material and temperature.\cite{Ashcroft} At low temperatures, long-wavelength (low-frequency) phonons dominate heat transport, whereas short-wavelength (high-frequency) phonons become more relevant at higher temperatures.Heat transport is primarily governed by acoustic phonons due to their higher group velocity compared to optical phonons. Phonon scattering can be enhanced by engineering the material’s micro- and nano-structure to create grain interfaces with spacings ranging from nanometers to micrometers. In fact, the characteristic size of the grain boundaries determines the surface roughness, and when the phonon wavelength becomes comparable to this dimension, diffuse scattering occurs (see Fig.~\ref{fig:surface_roughness}). In this process, phonons scatter randomly, losing memory of their initial direction. Notably, since the longest phonon wavelength is determined by the largest periodicity in the material, nanocrystalline materials composed entirely of nanocrystals do not benefit from microcrystalline phases, whereas microcrystalline materials can benefit from nano-inclusions. Since electrons have much shorter wavelengths than thermal phonons (based on the De Broglie relation and using the Fermi energy $E_F \sim 1-\SI{10}{eV}$ as kinetic energy, $\lambda = h / p < \SI{1}{nm}$), their transport is theoretically only minimally affected by grain boundaries in nanostructured materials.
\begin{figure}[H]
    \centering
    \includegraphics[width=0.5\linewidth]{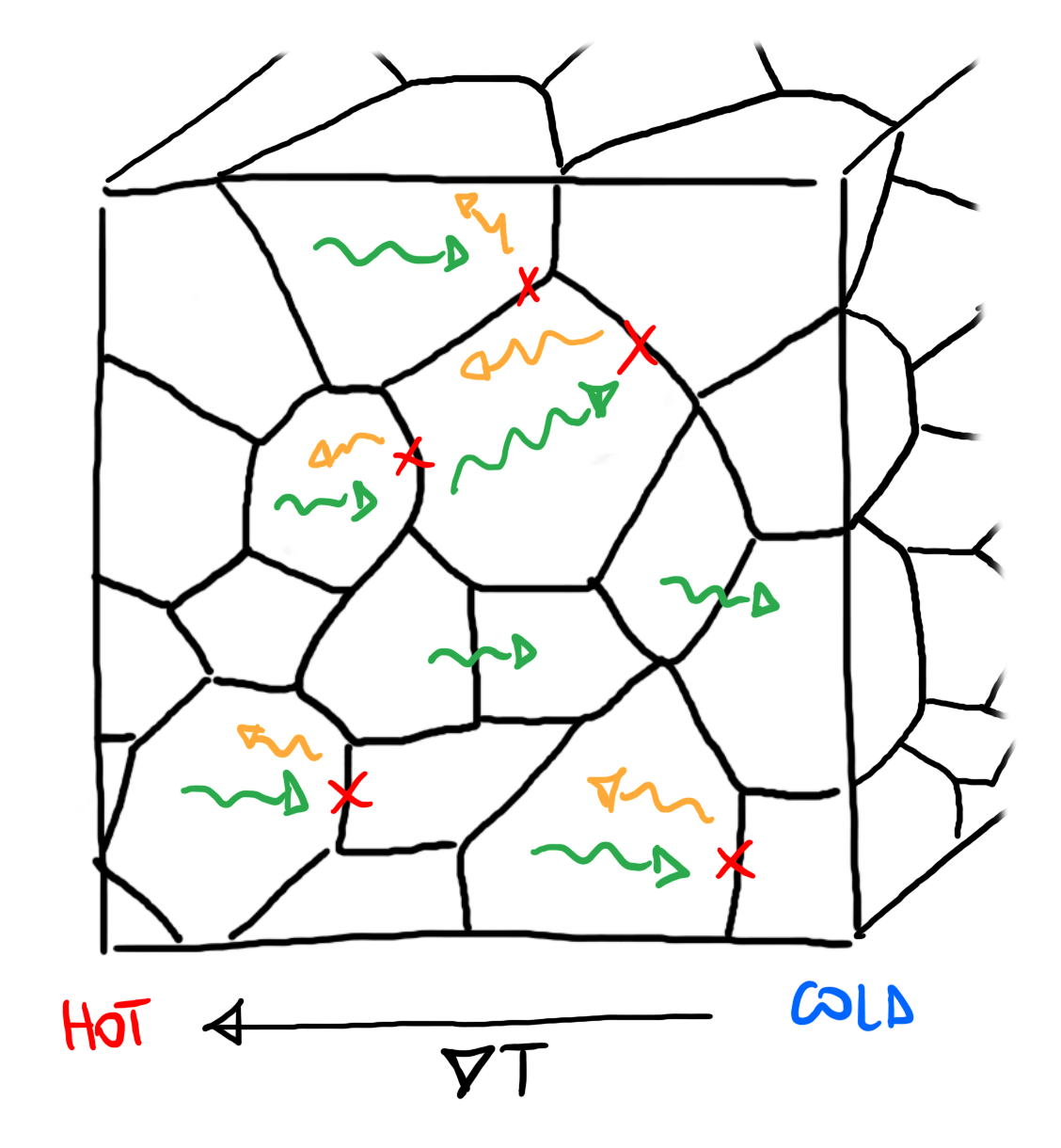}
    \caption{Illustration of a micro-structured solid, showing diffusion scattering of phonons at the grain boundaries. Given the wavelength $\lambda$ and the characteristic dimension $d$ of an obstacle, when $\lambda \gg d$, waves tend to bypass the obstacle with minimal disturbance. Conversely, when $\lambda \ll d$, waves behave particle-like and scatter less efficiently. The strongest scattering occurs when $\lambda \sim d$.}
    \label{fig:grains}
\end{figure} \noindent
\begin{figure}[H]
    \centering
    \includegraphics[width=1\linewidth]{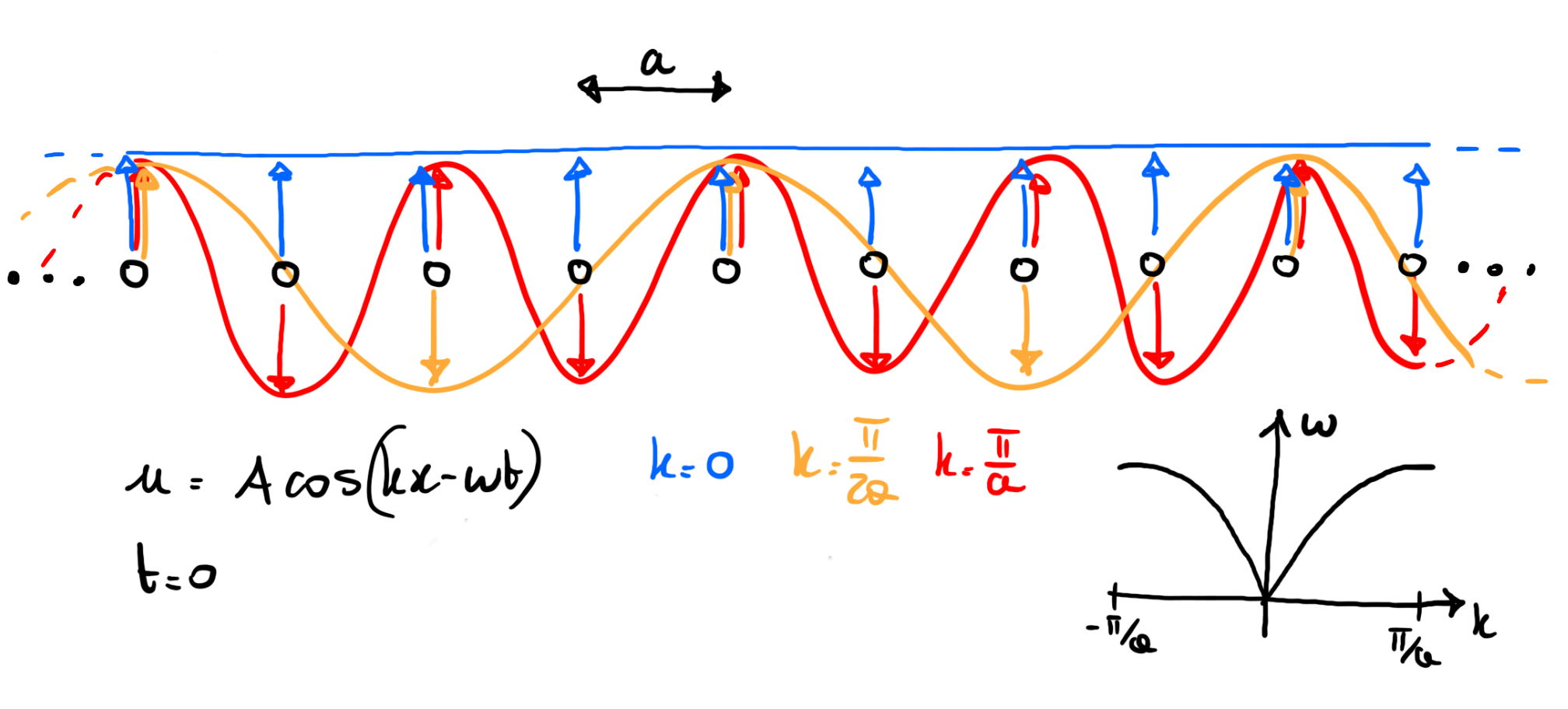}
    \caption{Illustration of three transversal normal modes of a linear mono-atomic chain, at time $t=0$, for $k=0$, $k=\pi/2 a$ and $k=\pi/a$, where $a$ is the lattice parameter. The longest wavelength corresponds to $\lambda = N a$, while the shortest to $2a$. In particular, the shortest wave (largest wave vector) is a standing wave.}
    \label{fig:normal_modes}
\end{figure} 
\noindent
In practice, micro- and nano-crystalline solids are synthesized via powder metallurgical techniques.\cite{K.Bux.B.Kaner.Chemical.Communications.2010, Dresselhaus.Gogna.Advanced.Materials.2007} Biswas et al. have demonstrated a record $zT \approx 2.2$ at circa \SI{900}{K} in p-type \ce{PbTe} through powder processing and spark plasma sintering, a technique that allows the formation of $\mu$m scale grains with nano-inclusions,\cite{Biswas.Kanatzidis.Nature.2012} while Ibàñez et al. have demonstrated a bottom-up approach to sinter solution-processed nanoparticles into n-type \ce{PbS}-\ce{Ag} pellets, with $zT$ reaching 1.7 at \SI{850}{K}.\cite{Ibáñez.Cabot.Nature.Communications.2016}  It is important to note that these results were obtained at medium to high temperatures, where the heat transport is dominated by phonons with short wavelengths. 

Two models are typically used to describe phonon scattering at grain boundaries: the Acoustic Mismatch Model (AMM) and the Diffusive Mismatch Model (DMM).\cite{little.canadian, Swartz.Pohl.Reviews.of.Modern.Physics.1989} Given the phonon wavelength $\lambda$ and the roughness scale $\delta$, the AMM applies to smooth surfaces, where $\lambda \gg \delta$ and phonons are reflected specularly. In contrast, the DMM is more suitable for rough surfaces, where $\lambda \ll \delta$ and phonons scatter diffusively. As such, the DMM is typically more appropriate to describe phonon scattering at grain boundaries in micro- and nano-crystalline materials. 
It is worth noting that, although the diffusive model was originally developed for interfaces between different materials, it also applies to interfaces between homogeneous materials, such as grain boundaries with different orientations. Grain misorientation creates boundaries that disrupt the crystal lattice periodicity, causing phonon scattering due to atomic mismatches. In addition to reducing $\kappa_l$, grain boundaries may introduce energetic barriers for electrons, effectively filtering out colder electrons from conduction. This increases the asymmetry of the free charge carrier distribution around the Fermi level, which should enhance the Seebeck coefficient while having a minimal impact on electrical conductivity, ultimately resulting in higher power factors $S^2 \sigma$.\cite{Narducci.Ottaviani.Journal.of.Solid.State.Chemistry.2012, Gayner.Amouyal.Advanced.Functional.Materials.2020} Therefore, materials characterized by grains should exhibit a much lower thermal conductivity than the pure crystal, and a similar electrical conductivity.
\begin{figure}[H]
    \centering
    \includegraphics[width=1\linewidth]{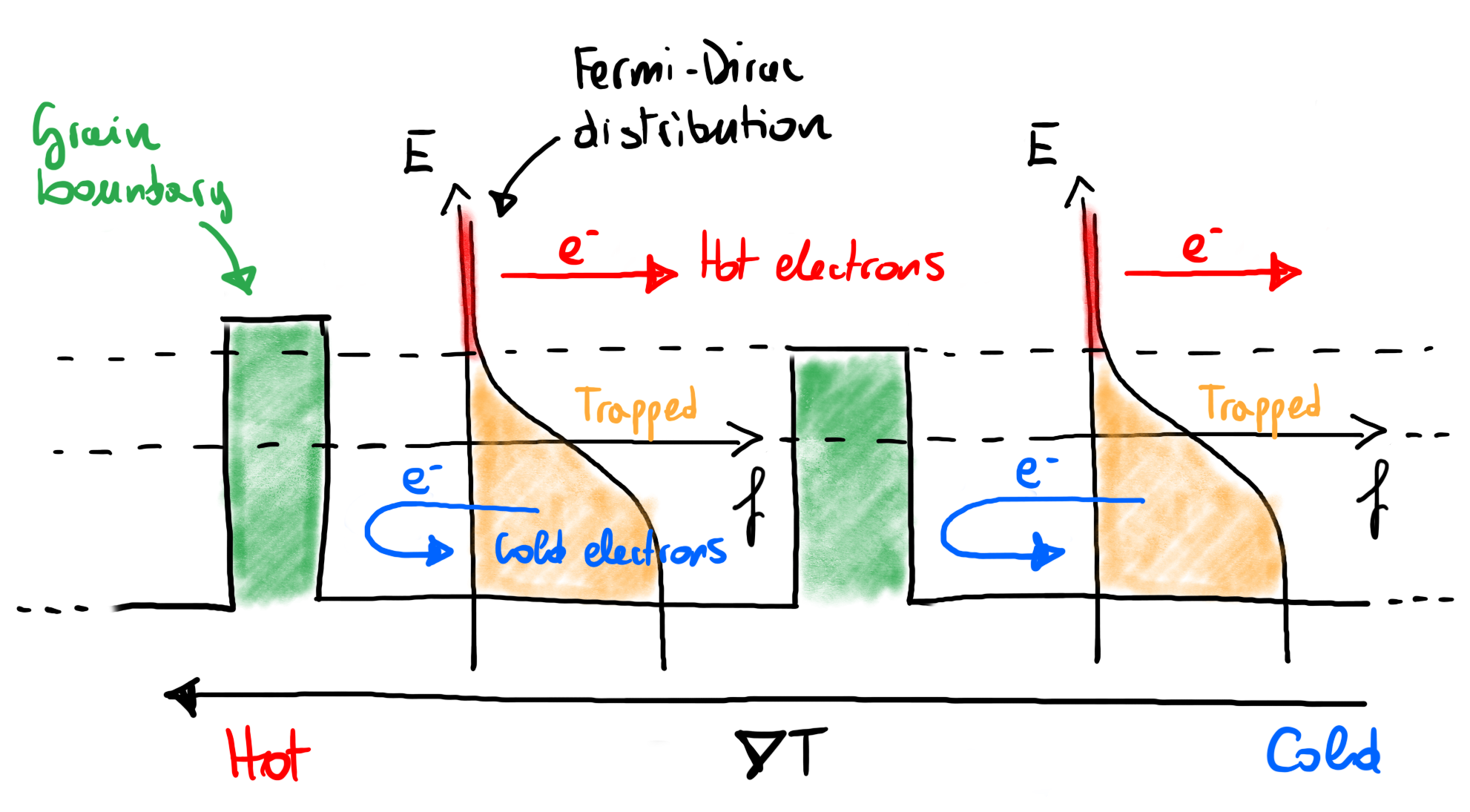}
    \caption{Illustration of the physics of energy filtering in a nano-crystalline solid. A simplified model showing energy barriers at the grain boundaries, where electrons are filtered based on their energy.}
    \label{fig:energy_filtering}
\end{figure}

\subsection{Low Dimensional Systems}
\begin{figure}[H]
    \centering
    \includegraphics[width=1\linewidth]{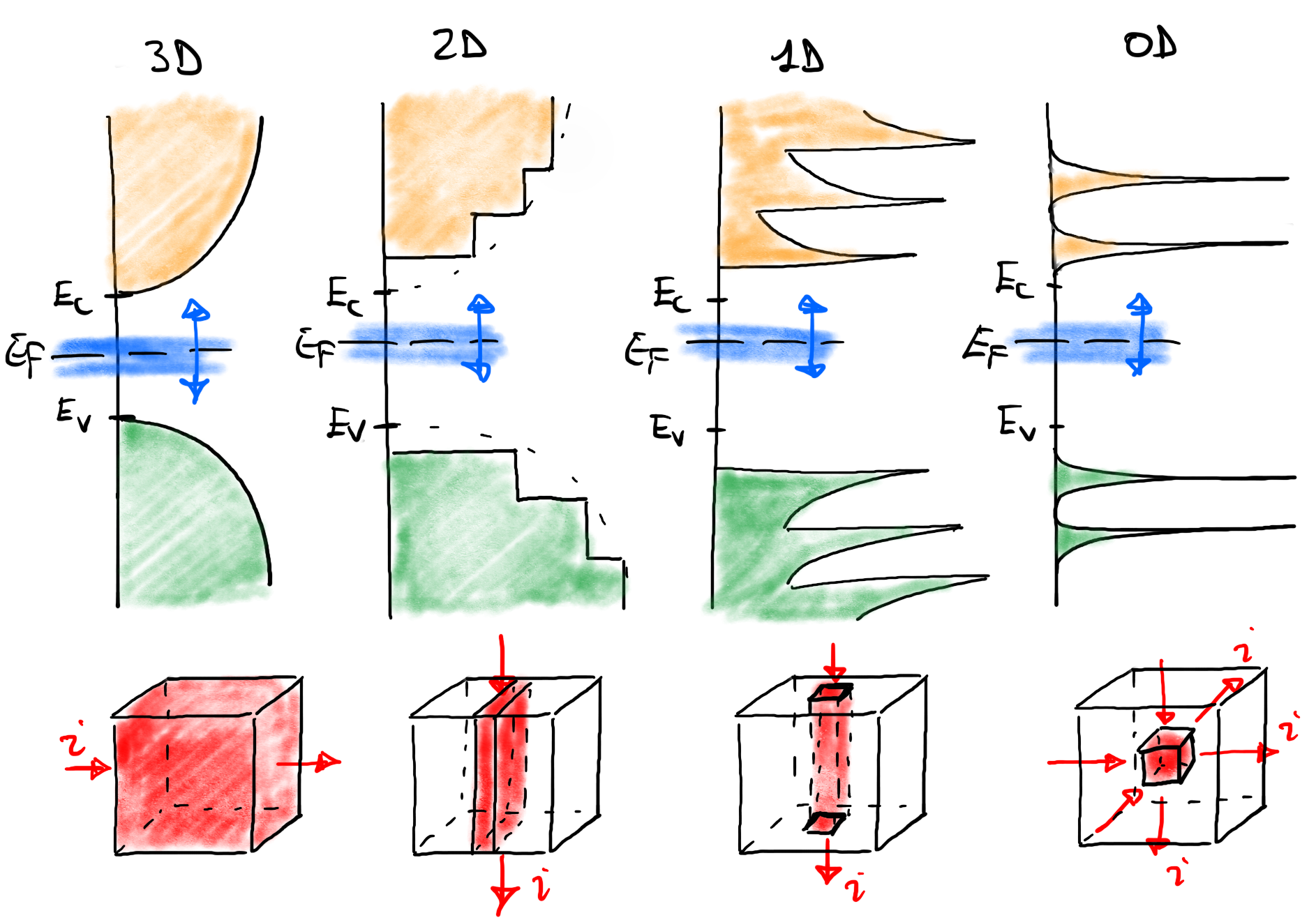}
    \caption{Qualitative representation of the density of states (DOS) in 3D, 2D, 1D, and 0D systems. Shifting the Fermi level can position it in a region of strong asymmetry within a few $k_B T$, which enhances the Seebeck coefficient. The red arrows represent the direction of the electron current.}
    \label{fig:quantum_dos}
\end{figure}

In two seminal papers from the 1990s, Hicks and Dresselhaus have theoretically shown that the $zT$ of quantum wells in multilayered superlattices and one-dimensional systems can be more than ten times larger than the $zT$ of their bulk counterparts when the charge and heat transport are perpendicular to the quantization axis.\cite{Hicks.Dresselhaus.Physical.Review.B.1993frn, Hicks.Dresselhaus.Physical.Review.B.1993} The enhancement of $zT$ would result from a combination of: (i) the electronic Density Of States (DOS) characteristic of quantum-confined systems (see Fig.\ref{fig:quantum_dos}), which may present a strong asymmetry around the Fermi level depending on doping, and (ii) the phonon scattering at the boundaries when the characteristic length of the nano-structure is smaller than the phonon mean free path $\lambda_{ph}$. 

Several experimental works on epitaxial multilayered superlattices have demonstrated remarkable thermoelectric properties of \textbf{Quantum Wells}, including \ce{PbTe/Pb_{1-x}Eu_{x}Te} with an estimated $zT$ of $2$ at \SI{300}{K},\cite{Hicks.Dresselhaus.Physical.Review.B.1996} and n-type \ce{PbTe/Te} with a $zT$ of $0.52$ at \SI{300}{K}.\cite{Harman.Walsh.Journal.of.Electronic.Materials.1999} However, their $zT$ remains to date far below the theoretical predictions due to many non-idealities that were not originally included in the theory, such as electron tunneling across- and/or thermal conductance along the superlattice barriers.\cite{Sofo.Mahan.Applied.Physics.Letters.1994} 
\noindent
\noindent

A way to improve the $zT$ of multilayered super-lattices consisted in incorporating a \textbf{Quantum Dot Superlattice} (QDSL) in between the superlattice layers, where a QDSL is a layer of quantum dots grown on a superlattice. With their highly asymmetric delta-like electron density of states, enhanced short-wavelength phonon scattering, and plausibly energy filtering effect, QDSL exhibited higher $zT$, with \ce{PbSe_{x}Te_{1-x}/PbTe} reaching $zT=0.9$ at \SI{300}{K}.\cite{Harman.Walsh.Journal.of.Electronic.Materials.2000} However, the $zT$ enhancement were only marginal at the expense of quite involved fabrication processes. 

Many experimental studies on \textbf{quantum wires} have confirmed their superior thermoelectric properties, primarily due to their extraordinarily low thermal conductivity, which results from roughness-dependent phonon scattering at the surface and the flattening of the dispersion relation due to quantum confinement.\cite{Wingert.Chen.Nano.Letters.2015, Lim.Yang.Nano.Letters.2012} Si nano-wires exhibit a $zT$ of $0.6$ at \SI{300}{K},\cite{Hochbaum.Yang.Nature.2008, Pennelli.Pennelli.Beilstein.Journal.of.Nanotechnology.2014} i.e. more than $100$ times larger than the $zT$ of bulk Si, with a thermal conductivity as low as \SI{2}{\watt\per\meter\per\kelvin} against the typical $\sim\SI{150}{\watt\per\meter\per\kelvin}$ of the bulk. Si has also been shown to retain a very low thermal conductivity of \SI{4.6}{\watt\per\meter\per\kelvin} when grown in very dense vertical nano-wire arrays,\cite{Pennelli.Dimaggio.Nanotechnology.2018} and $\sim \SI{20}{\watt\per\meter\per\kelvin}$ when grown in thin membranes or nano-ribbons.\cite{Pennelli.Macucci.IEEE.Transactions.on.Nanotechnology.2018} This has enabled single-leg TEGs based on Si nanowire arrays with a power output up to $\approx \SI{50}{\micro\watt\per\centi\meter\squared}$ for $\Delta T \approx \SI{10}{K}$ at room temperature.\cite{Pennelli.Bruschi.Nano.Letters.2013} p-type and n-type \ce{Si_{1-x}Ge_{x}} alloy nanowires were reported to have a $zT$ of $\approx 0.18$ and $\approx0.2$ at \SI{300}{K},\cite{Martinez.Swartzentruber.Journal.of.Applied.Physics.2011, Lee.Kim.Nano.Letters.2012} respectively, also thanks to an extremely low thermal conductivity of $\approx \SI{1.1}{\watt\per\meter\per\kelvin}$,\cite{Lee.Kim.Nano.Letters.2012} and TEGs based on \ce{SiGe} nanowires planar arrays were also demonstrated with a power output of $\approx \SI{10}{\micro\watt\per\centi\meter\squared}$ for $\Delta T \sim \SI{200}{K}$.\cite{Noyan.Fonseca.Nano.Energy.2019} It is now clear that the reduction of $\kappa_l$ due to phonon-surface scattering is marginal in all materials characterized by a small phonon mean free path, such as the Chalcogenides of Bismuth.\cite{Chen.Li.Chemical.Reviews.2019} On the other hand, although the $zT$ of nanowires at \SI{300}{K} does not exceed one,\cite{Chen.Li.Chemical.Reviews.2019} nanowires have enabled room temperature thermoelectric applications for CMOS-technology-compatible materials like \ce{Si} and \ce{SiGe} alloys, with obvious benefits coming from fabrication processes that are well established in the semiconductor industry. 

\textbf{2D Materials} represent the lower-dimensional limit of quantum wells, being essentially entirely surface-based. However, the Density of States (DOS) of 2D materials differs significantly from that of quantum wells. For ideal quantum wells, the DOS is constant near the band edges due to their parabolic energy dispersion (see Fig.~\ref{fig:quantum_dos}). In contrast, 2D materials like graphene and Transition Metal Dichalcogenides (TMDs) exhibit distinct behaviors due to their unique band structures. Graphene, in particular, is gapless and exhibits extraordinary thermal conductivity of up to \SI{5000}{\watt\per\meter\per\kelvin}, owing to high phonon group velocities (enabled by strong bonds and low atomic mass) and long phonon mean free paths (due to minimal defect scattering and weak anharmonicity), making it unsuitable for thermoelectric energy conversion.\cite{Zuev.Kim.Physical.Review.Letters.2009, Ghosh.Balandin.Nature.Materials.2010} On the other hand, TMDs such as \ce{MoS2} and \ce{SnS2} are semiconducting and have lower thermal conductivity (\(\kappa \sim \SI{10}{}-\SI{50}{\watt\per\meter\per\kelvin}\)), resulting in $zT \approx 0.1$ at room temperature.\cite{Pallecchi.Marré.Nano.Futures.2020}
Notably, as 2D materials are entirely surface-based, their transport properties—and therefore their thermoelectric performance—are strongly influenced by interactions with their substrates and the charge carrier injection mechanism. 
To enhance the $zT$ of 2D materials, novel nanostructuring techniques are being explored, along with the effects of the number of atomic layers and strain, which modify their energy dispersion relation.\cite{Markov.Zebarjadi.Nanoscale.and.Microscale.Thermophysical.Engineering.2019} 

Nanoribbons closely resemble \textbf{1D Material}s. Although experimental data on these materials are currently limited, theoretical calculations predict $zT > 3$ in nanoribbons of graphene, attributed to the energy gap induced by lateral confinement and phonon boundary scattering.\cite{Sevinçli.Cuniberti.Physical.Review.B.2010}

Finally, \textbf{Quantum Dots (QDs)}—0D systems where electrons are confined in all three spatial dimensions—exhibit discrete energy levels (see Fig.~\ref{fig:quantum_dos}).Theoretically, single-level quantum dots have been proposed as ideal Carnot cycle engines,\cite{Humphrey.Linke.Physical.Review.Letters.2002, Esposito.Broeck.EPL.Europhysics.Letters.2009, Nakpathomkun.Linke.Physical.Review.B.2010} and experimental studies have shown that \ce{InP} quantum dots embedded in \ce{InAs} semiconductor nanowires can achieve efficiencies close to the Curzon-Ahlborn limit at maximum power, owing to a combination of energy filtering and phonon scattering.\cite{Josefsson.Linke.Nature.Nanotechnology.2018} However, these experimental results are likely overestimated, as the heat flux in these experiments was theoretically calculated without accounting for phonon contributions. To block the phonon contribution to the heat flux, theoretical three-terminal particle-exchange devices—comprising two cold electrodes and one electrically insulated hot reservoir—have been proposed.\cite{Sothmann.Jordan.Nanotechnology.2014} Experimental demonstrations have been conducted based on either two capacitively coupled QDs or two QDs in series defined in a high-mobility two-dimensional electron gas.\cite{Thierschmann.Molenkamp.Nature.Nanotechnology.2015, Jaliel.Smith.Physical.Review.Letters.2019} Unfortunately, these devices must operate in the \SI{}{\milli\kelvin} temperature range to ensure that the thermal energy, $k_B T$, is much smaller than the quantum dot energy level spacing. 
As small molecules can provide a larger energy level spacing between their Highest Occupied Molecular Orbital (HOMO) and Lowest Unoccupied Molecular Orbital (LUMO), they are being studied for room-temperature thermoelectricity. While most research is fundamental and uses Scanning Tunneling Microscope setups, where the molecule is temporarily trapped between the substrate and the conducting tip,\cite{Malen.Segalman.Chemical.Physics.Letters.2010, Gemma.Gotsmann.Nature.Communications.2023} molecular solid-state devices have been demonstrated with an estimated record $zT$ of 0.7 at \SI{3}{\kelvin} for \ce{Au-[Gd(tpy-SH)2(NCS)3]-Au} junctions.\cite{Kim.Reddy.Nature.Nanotechnology.2014, Gehring.Mol.Nano.Letters.2017, Gehring.Zant.Nature.Nanotechnology.2021} In these systems, transport properties are typically calculated using the Landauer formalism or by solving the full quantum mechanical problem with an Hamiltonian for the open system (Green's functions).

A critical challenge for low-dimensional systems, and especially 2D, 1D materials and QDs, is chemical doping due to their reduced dimensions and surface-dominated nature. As a result, these systems are often integrated into field-effect configurations, where electrostatic gating and/or a strong bias voltage allow control of the carrier concentration, enabling their thermoelectric performance to be optimized dynamically. Additionally, some of these systems must operate at very low temperature to preserve their quantum properties. Therefore, while these systems are essential for fundamental studies, to date, their practical applications remain elusive.
\begin{figure}[H]
    \includegraphics[width=1\linewidth]{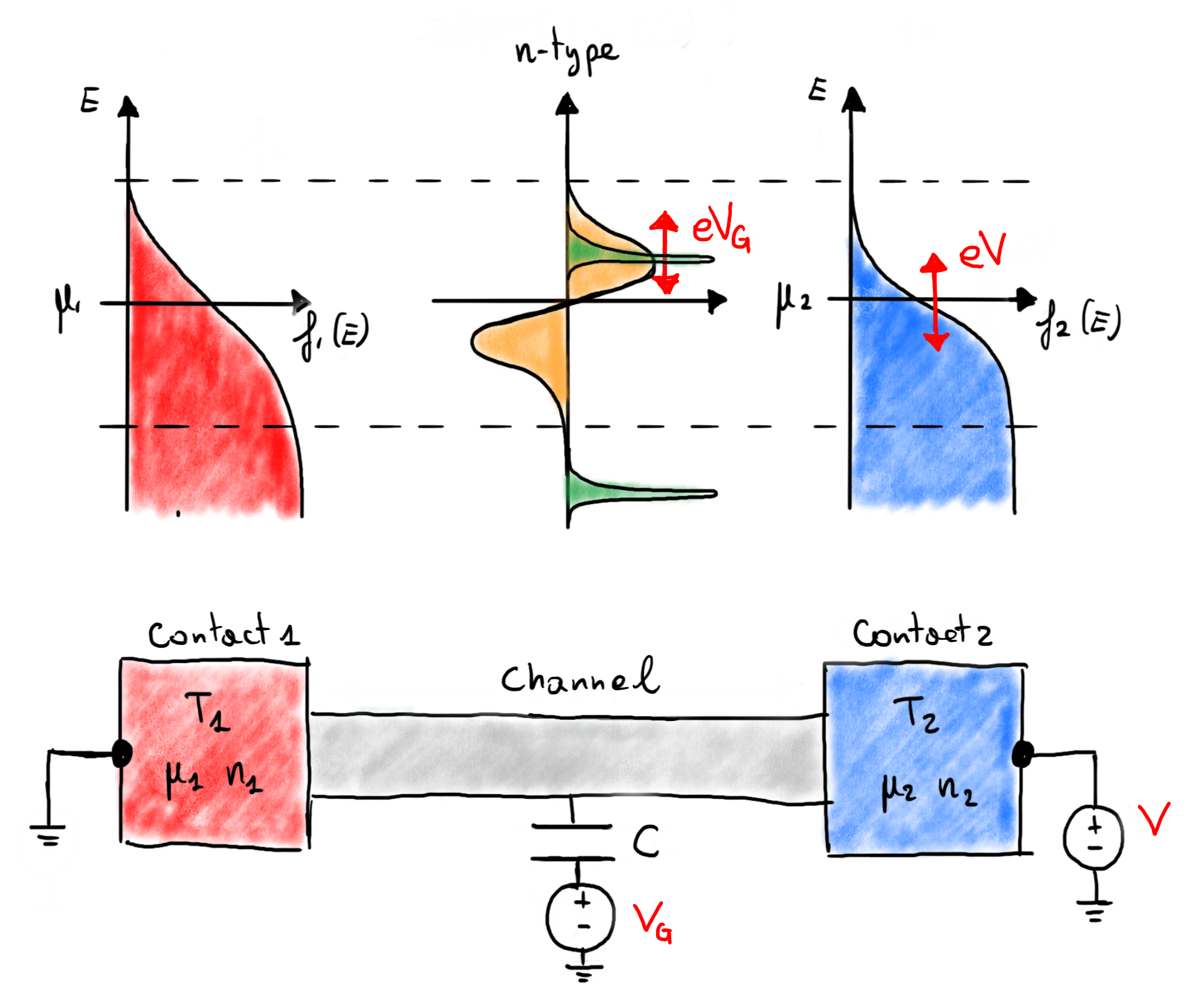}
    \caption{Illustration of the physical principles behind a nanoscale junction in a field effect configuration, where the channel exhibits a discrete-like DOS. By gating, the DOS can be shifted in energy, so that only electrons above or below the Fermi level conduct.}
    \label{fig:nanoscale_interfaces}
\end{figure}\noindent

\subsection{Phononic Crystals}\label{sec:phononic_crystals}
\begin{figure}[H]
    \centering
    \includegraphics[width=1\linewidth]{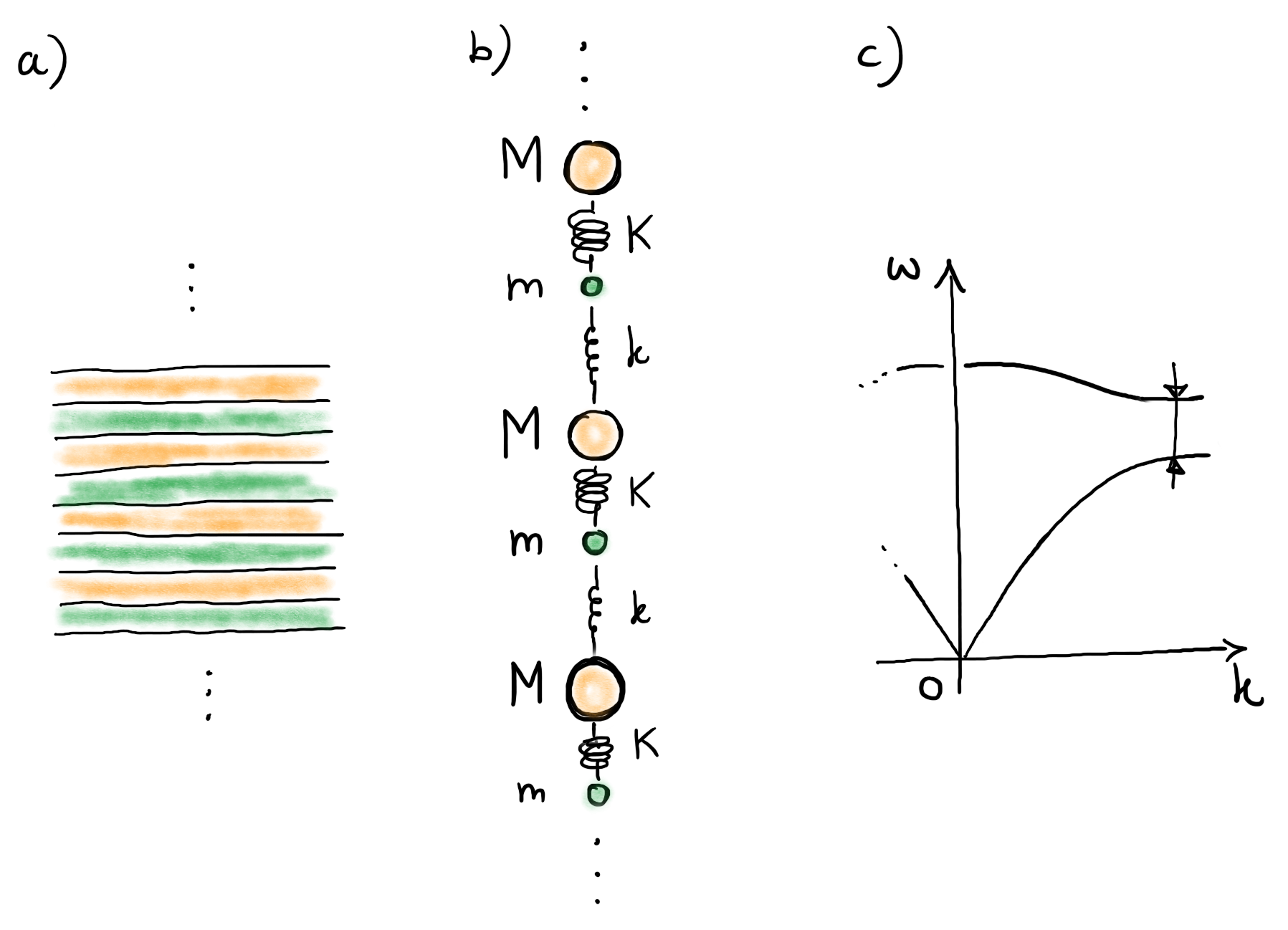}
    \caption{(a) Illustration of a typical 1D phononic nanocrystal made of a period heterostructure. (b) A simple model to describe the system, and (c) its corresponding dispersion relation.}
    \label{fig:phononic_nanocrystals}
\end{figure}

Phononic crystals are 1D, 2D or 3D periodic nanostructures with periodicity on the order of the phonon wavelength (typically a few nm).\cite{Nomura.Volz.Materials.Today.Physics.2022} These materials can be engineered to exhibit very low thermal conductivity by combining two types of phonon scattering mechanisms: coherent (or Bragg) scattering and incoherent (or diffusive) boundary scattering. In coherent scattering, phonons are reflected due to the periodic structure, similar to how light is diffracted in a crystal. In contrast, in incoherent scattering phonons lose their directional coherence and scatter randomly. Coherent scattering occurs when the phonon wavelength is comparable to the characteristic length of the periodic structure, while diffusive scattering, typically described by the Diffusion Mismatch Model (DMM) mentioned in Section \ref{sec:micro_nano_grains}, occurs when the phonon mean free path is on the same order as the interface roughness.\cite{Hussein.Honarvar.Advanced.Functional.Materials.2020, Alaie.El-Kady.Nature.Communications.2015} Importantly, as predicted by the Bloch theorem for periodic structures, coherent scattering can result in phononic band gaps, further reducing the thermal conductivity of the sample due to a lack of available phonon modes and reduced group velocity.\cite{Jiang.Rabczuk.Nano.Letters.2013, Zhang.Li.Nano.Letters.2017, Yu.Heath.Nature.Nanotechnology.2010} Additionally, coherent scattering can be engineered to fabricate heat guides, such as heat lenses.\cite{Anufriev.Nomura.Nature.Communications.2017} 
The typical \textbf{1D phononic crystal} is a superlattice of periodically stacked heterostructures, where the temperature gradient—and thus the charge and heat transport—are orthogonal to the heterostructure growth direction. In such systems, heat transport is limited by phonon scattering at the superlattice interfaces due to lattice mismatch imperfections (incoherent or diffusive scattering) and by acoustic impedance mismatch across heterogeneous interfaces, which result in the reflection of low-frequency phonons (coherent scattering).\cite{Chen.Chen.Physical.Review.B.1998, Venkatasubramanian.Venkatasubramanian.Semiconductors.and.Semimetals.2001, Cahill.Phillpot.Journal.of.Applied.Physics.2003, Luckyanova.Chen.Science.2012} In these systems, energy filtering may also improve the Seebeck coefficient.\cite{Thesberg.Neophytou.Journal.of.Applied.Physics.2016} A proper design of the superlattice could also result in acoustic phonon band gaps.\cite{Harman.LaForge.Science.2002, Narayanamurti.Wiegmann.Physical.Review.Letters.1979, Simkin.Mahan.Physical.Review.Letters.2000} To visualize the physics behind 1D phononic crystals, one can refer to Fig. \ref{fig:phononic_nanocrystals}: here, a linear chain made of two different masses and spring constants plays the role of the periodic structure. It is well-known and widely described in the literature that such a system results in a phonon dispersion relation with a phononic band gap at the Brillouin zone edges.\cite{Ashcroft} 
 \noindent
In 2001, Venkatasubramanian et al. measured extraordinary values of $zT=2.4$ at \SI{300}{K} for the p-type \ce{Bi2Te3}/\ce{Sb2Te3} superlattice and $zT=1.4$ at \SI{300}{K} for the n-type \ce{Bi2Te3}/\ce{Bi2Te_{2.83}Se_{0.17}} superlattice.\cite{Venkatasubramanian.O'Quinn.Nature.2001} These are the highest $zT$ values ever measured in quantum systems at room temperature.
The most common \textbf{2D phononic crystals} are suspended membranes with a periodic array of holes or inclusions. Tang et al. demonstrated experimental values of thermal conductivity ($\kappa$) reaching down to $\SI{1}{-}\SI{2}{\watt\per\meter\per\kelvin}$ and a $zT$ of approximately 0.4 at room temperature in \SI{100}{\nano\meter}-thick Si membranes with nm-scale air-filled holes and a \SI{55}{\nano\meter} pitch.\cite{Tang.Yang.Nano.Letters.2010}
\textbf{3D phononic crystals}, such as nm-scale sphere-filled solids, are still largely theoretical due to the complexity involved in their fabrication, despite their promising potential.

\subsection{Phononic Metamaterials}
Phononic metamaterials are nanostructured materials that incorporate periodically arranged resonators—structural features that support localized vibrational modes at specific frequencies. These resonators often take the form of ad-structures, such as nanoscale pillars, suspended membranes, or inclusions embedded within the bulk material. Unlike phononic crystals (see Section \ref{sec:phononic_crystals}), which typically involve periodic modifications such as holes or inclusions carved directly within the material, phononic metamaterials emphasize the addition of structural elements (ad-structures) on top of the material to manipulate vibrational modes. This distinction reflects the different mechanisms by which phonons interact with the periodic structures in each system.
\begin{figure}[H]
    \centering
    \includegraphics[width=0.9\linewidth]{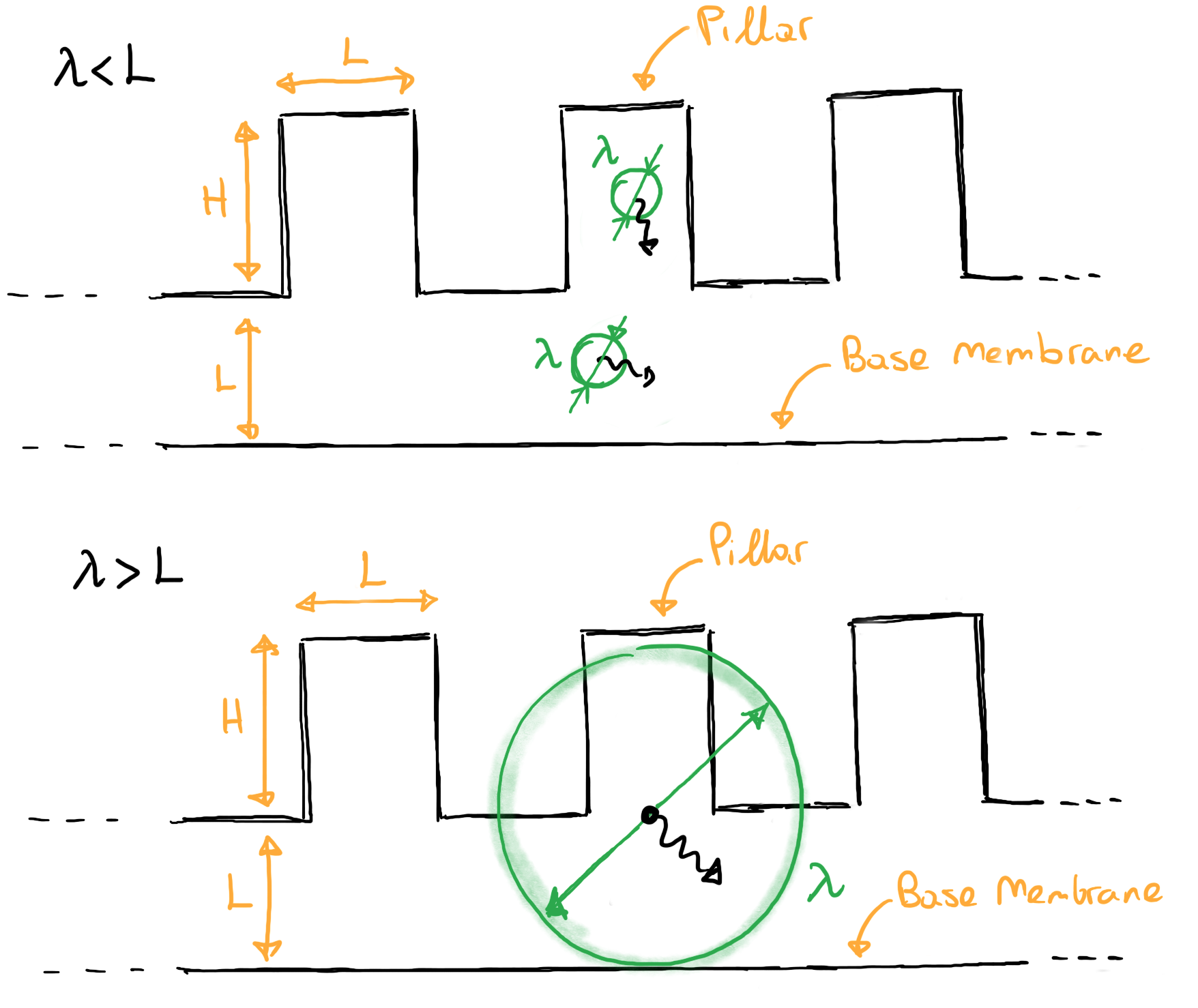}
    \caption{Illustration of phononic metamaterials incorporating ad-structures in the form of pillars. Reproduced from Hussein et al.\cite{Hussein.Honarvar.Advanced.Functional.Materials.2020}}
    \label{fig:phononic_metamaterials}
\end{figure} \noindent
In particular, the interaction between the phonon modes of these resonators and the phonon modes of the bulk material leads to band flattening, and thus reduced phonon group velocity and lower lattice thermal conductivity, and, in some cases, the opening of phononic band gaps.\cite{Davis.Hussein.Physical.Review.Letters.2014, Hussein.Honarvar.Advanced.Functional.Materials.2020} This phenomenon, known as \textit{avoided crossing} and previously discussed in Section \ref{sec:lattice_disorder}, occurs naturally in materials where atoms are encapsulated in caged structures, such as Clathrates.\cite{Beretta.Caironi.Materials.Science.and.Engineering.R.Reports.2019} However, in Clathrates, the hybridization is typically limited to the modes of the guest atoms. In contrast, phononic metamaterials can be designed to enable hybridization across the entire acoustic phonon spectrum. This offers the advantage of controlling the phonon velocity by a proper choice of resonators. The great advantage of phononic metamaterials for thermoelectrics resides in their ability to manipulate the lattice thermal conductivity $\kappa_l$ via these artificial ad-structures without significantly affecting the electrical conductivity $\sigma$. 
While most research on phononic metamaterials remains theoretical, recent experimental studies have demonstrated thermal conductivities of $\kappa \approx 30 - \SI{50}{\watt\per\meter\per\kelvin}$ in Si membranes with $\mu$m-scale thick \ce{GaN} nano-pillars and in all-\ce{Si} suspended wires with nano-walls.\cite{Maire.Nomura.Scientific.Reports.2018, Spann.Bertness.Advanced.Materials.2023} However, a comprehensive thermoelectric characterization of phononic \textit{thermoelectric} metamaterials has yet to be done.

\subsection{Thermionic Generation}
Proposed by Shakouri and Mahan in the late 1990s,\cite{Shakouri.Bowers.Applied.Physics.Letters.1997it, Mahan.Mahan.Journal.of.Applied.Physics.1994} and inspired by the vacuum-diode-based Thermo-Electron Engines of the 1950s,\cite{Hatsopoulos.Kaye.Journal.of.Applied.Physics.1958, Houston.Houston.Journal.of.Applied.Physics.1959} these systems consist of a reversed-biased Metal-Semiconductor-Metal (MSM) heterostructure, with a temperature difference $\Delta T$ established between the two metal electrodes. Assuming that the electron transport is ballistic, the energy barriers between the metals and the semiconductor filter out the cold electrons, resulting in a high Seebeck coefficient and power factor. Due to the thin nature of the heterostructure, several MSM heterostructures in series are required to minimize the lattice thermal conductance and achieve a theoretical efficiency of up to \SI{80}{\percent} of the Carnot cycle, which is twice as that of a conventional thermoelectric generator.\cite{Mahan.Bartkowiak.Journal.of.Applied.Physics.1998} 
\begin{figure}[H]
    \centering
    \includegraphics[width=0.9\linewidth]{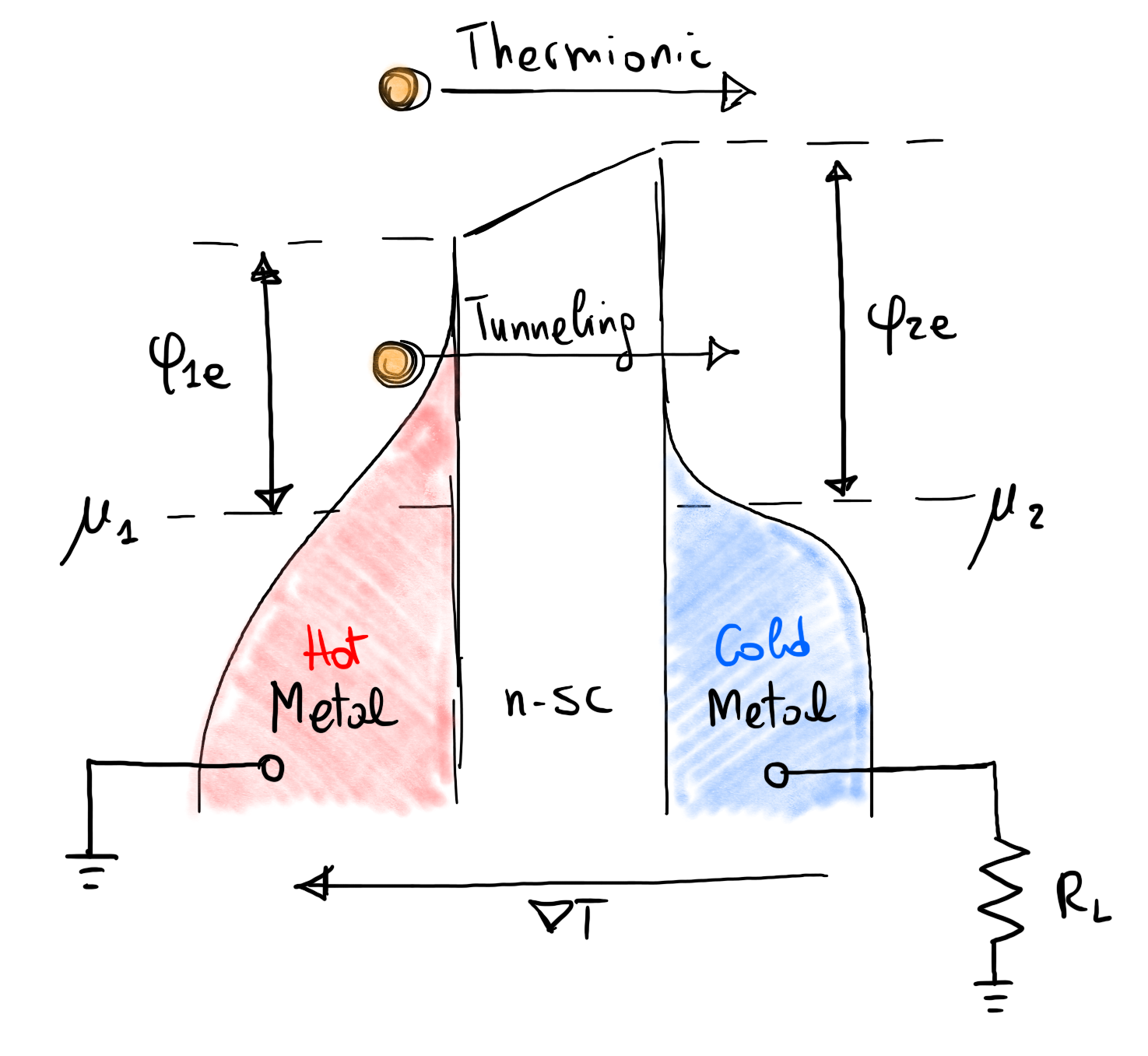}
    \caption{Illustration of a single stage thermionic generation operating under a temperature gradient, without external bias.}
    \label{fig:thermionic}
\end{figure}\noindent
Recently, Liang et al. extended the thermionic energy conversion theory to van der Waals (vdW) heterostructures in the linear regime,\cite{Liang.Ang.Scientific.Reports.2017} and their theory confirms that thermionic energy converters can outperform conventional thermoelectric generators. While not explicitly discussed by the authors, it is interesting to note that a single vdW heterostructure may suffice for a thermoelectric generator, owing to the strong interfacial heat resistance resulting from the spacing between vdW layers, which leads to very low lattice thermal conductivity.  Despite the promising potential of thermionic generation in vdW heterostructures, only a few theoretical studies have been conducted on thermoelectric transport in vertical vdW stacks, such as Gr/\ce{MoS_2}/Gr and Gr/C60/Gr heterostructures,\cite{Sadeghi.Lambert.2D.Materials.2016} and very few experimental studies, including Gr/h-BN/Gr and Au/h-BN/Gr heterostructures,\cite{Chen.Cronin.Nano.Research.2015, Poudel.Cronin.Scientific.Reports.2017} primarily due to challenges in fabricating interfaces with appropriate energy barriers and in characterizing thermoelectric transport.

\section{Perspectives}
The strategies for enhancing the thermoelectric figure of merit $zT$ are rarely pursued in isolation. Instead, the most promising approaches often arise from the combination of multiple techniques. For example, tailoring grain size in micro- and nano-crystalline solids together with optimizing chemical doping are strategies typically adopted together. Similarly, combining doping optimization with the use of low-dimensional systems such as quantum wells, nanoribbons, or quantum dots opens new pathways for tuning energy filtering and thus carrier transport. 
Low-dimensional systems, in particular, hold great promise and continue to be a focus of intense investigation due to their unique quantum properties. While they pave the way for potentially more efficient thermoelectric devices, their practical implementation is often limited by intrinsic constraints. For instance, quantum dots, while theoretically ideal for energy filtering, cannot sustain large current densities due to their discrete energy levels and confinement effects, and must be operated at very low temperature. This highlights the critical challenge of preserving quantum properties while also exploiting them at a macroscopic scale, in a temperature range of interest for thermoelectric generation, to achieve meaningful power outputs, open circuit voltage and short circuit current. 
To address these limitations, low-dimensional systems may need to be integrated into larger-scale, microscopically engineered systems, such as those found in micro- and nano-crystalline solids. Such integration could enable the observation and utilization of quantum properties at a macroscopic level, bridging the gap between fundamental studies and real-world applications.

\section*{Acknowledgments}
None to report.

\bibliographystyle{unsrt}
\bibliography{main}

\end{multicols}
\end{document}